\documentclass[showpacs,preprintnumbers,prb,superscriptaddress]{revtex4}
\usepackage{graphicx}

\begin{document}

\title{Adaptive quantum design of atomic clusters}
\author{Jason Thalken}
\affiliation{Department of Physics and Astronomy, University of Southern California, Los
Angeles, CA 90089-0484}
\author{Yu Chen}
\affiliation{Department of Electrical Engineering, University of Southern California, Los
Angeles, CA 90089-2533}
\author{A.F.J. Levi}
\affiliation{Department of Physics and Astronomy, University of Southern California, Los
Angeles, CA 90089-0484}
\affiliation{Department of Electrical Engineering, University of Southern California, Los
Angeles, CA 90089-2533}
\author{Stephan Haas}
\affiliation{Department of Physics and Astronomy, University of Southern California, Los
Angeles, CA 90089-0484}
\date{\today}
\pacs{PACS numbers: }

\begin{abstract}
Adaptive quantum design identifies the best broken-symmetry configurations
of atoms and molecules that enable a desired target function response. In
this work, numerical optimization is used to design atomic clusters with
specified quasiparticle densities of states. The dominant self-assembled
building blocks of these engineered quantum systems are found to depend on
the symmetry of the target function. For example, particle-hole symmetric
spectra can be constructed from a dilute configuration of atomic dimers,
whereas more complex structures such as trimers and quadrumers are required
for asymmetric target functions. The convergence of the optimization
algorithms depends on the shape of the target function, the density of
atoms, and constraints due to substrates and boundary conditions. Hybrids of
steepest descent methods, simulated annealing, and genetic algorithms are
found to be most efficient.
\end{abstract}

\maketitle

\section{Introduction}

In the near future, it will likely be possible to control the
precise spatial positions of atoms and molecules using the
experimental techniques now being developed by nanoscience. To
complement these emerging capabilities it is clear that a new set
of theoretical tools has to be developed to assist in the
exploration of a potentially vast number of atom configurations
and a corresponding enormous range of physical properties. In this
paper, we outline an approach, which we call ``adaptive quantum
design", that sets out to address this challenging task. In
contrast to classical systems, atomic scale devices exhibit
quantum fluctuations and collective quantum phenomena caused by
particle interactions. Besides offering an excellent testing
ground for models of correlated electrons, they also force us to
reconsider conventional paradigms of condensed matter physics,
such as crystal symmetries that are imposed by nature. In some
instances, such symmetries need to be explicitly broken in order
to enable or optimize a desired system response. Consider for
example the quasiparticle density of states in tight-binding
systems, which is the subject of the present study. In
translationally invariant structures, i.e. crystals, it is well
known that the spectral response function exhibits van-Hove
singularities at positions of low dispersion, such as the band
edges in a one-dimensional chain or the band center in a
two-dimensional square lattice. These enhancements of the density
of states can be very useful in amplifying system responses such
as optical conductivity at specific incident energies. It is
therefore important to be able to control the positions and shapes
of such features by adaptive design techniques applied to models
which capture the essential degrees of freedom of interacting
atomic clusters.

Traditional ad-hoc methods for the design of nanoscale devices will likely
miss many possible configurations. At the same time, it is unrealistic to
expect individuals to manually explore the vast phase space of possibilities
for a particular device function. The proposed solution to this difficult
design problem is to employ machine-based searches of configuration space
that enable user-defined target functions. Adaptive quantum design solves
the inverse problem by numerically identifying the best broken-symmetry
spatial configuration of atoms and molecules that produce a desired target
function response.

The two major ingredients of adaptive quantum design are the physical model,
which in this work evaluates the electronic density of states for a
particular spatial arrangement of the atoms, and the search algorithm that
finds the global minimum in the parameter space of all possible
configurations. This problem is typically highly underdetermined in the
sense that there can be several atomic configurations that yield a system
response very close to the desired target function. Often, the associated
landscape of solutions is shallow and has many nearly degenerate local
minima.

Adaptive quantum design therefore relies heavily on efficient algorithms
that accurately model the interacting system and find its optimal
configuration, i.e. the global minimum in the available parameter space that
yields the best match to the desired target function. This target response
may be a specific angular transmission of a photonic crystal, a fine-tuned
Josephson current along a superconducting junction array, or an
energy-dependent density of states profile of an atomic cluster - the case
on which we focus here. In this work, various numerical search tools are
applied, including guided random walk, simulated and triggered annealing,
and genetic algorithms. In particular the last set of techniques is most
efficiently implemented on parallel computers. It is observed that in
practice hybrids of these methods yield the best results, i.e. fast
convergence towards a global minimum.

The particular example of a long-range tight-binding model is
chosen for this study because it captures essential features of
correlated quantum mechanical systems, and yet permits fast
numerical diagonalization of relatively large clusters with broken
translational symmetry. In a typical optimization run, these
``function calls" occur 100-10,000 times. This model is therefore
suitable for developing and testing adaptive design algorithms,
and should be viewed as an initial step towards the design of
atomic clusters. The goal of this work is to test the
applicability and limits of adaptive design techniques on a simple
but non-trivial quantum system.

\section{Long-range tight-binding model}

The tight-binding approach is an effective tool to describe the band
structure of electronic systems. It is commonly used to model the relevant
bands close to the Fermi level, obtained from complex density functional
theory calculations. Since in this work symmetry-breaking, non-periodic
configurations are considered, a long-range variant of the tight-binding
model with overlap integrals depending on the variable inter-atomic distance
has to be used. Its Hamiltonian is given by
\begin{eqnarray}
H = - \sum_{i,j} t_{i,j} \left( c^{\dagger}_i c_j + c_i c^{\dagger}_j
\right),
\end{eqnarray}
where $c^{\dagger}_i$ and $c_i$ are creation and annihilation operators at a
site $\vec{r}_i$, and the sum over pairs of atoms is restricted to avoid
double-counting. Here the spatial decay of the overlap integral $t_{i,j}$
is parametrized by a power-law,
\begin{eqnarray}
t_{i,j} = \frac{t}{|\vec{r}_i - \vec{r}_j |^{\alpha} },
\end{eqnarray}
where the exponent $\alpha$ is taken to be 3.0 throughout the paper unless
mentioned otherwise. \cite{harrison} This parametrization reflects an
algebraic variation of the overlap integral with inter-atomic separation.
The choice of sign in the Hamiltonian follows the convention for s-orbitals.
However, this simple implementation of the tight-binding model does not
account for the orbital directionality of realistic Au, Ag, or Pd atomic
wave functions. More sophisticated and numerically expensive techniques,
such as the local density approximation, would be needed to make
quantitative predictions for these systems.

\begin{figure}[h]
\includegraphics[width=6.5cm]{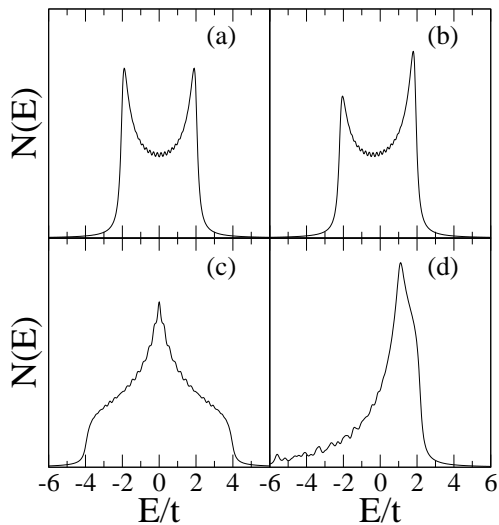}
\caption{ Density of states in spatially invariant tight-binding systems
with periodic boundary conditions. (a) Nearest-neighbor chain (no long-range
overlap integral) with 30 atoms. (b) Same as (a), but with long-range
overlaps according to Eq. 2. (c) Nearest-neighbor square lattice with 400
atoms. (d) Same as (c), but with long-range overlaps. }
\end{figure}

The Hamiltonian matrix of the long-range tight-binding model in the basis of
single-particle states is non-sparse, and only its diagonal matrix elements
vanish. In order to obtain the spectrum, the matrix is diagonalized
numerically for finite clusters. In Fig. 1, the resulting densities of
states of translationally invariant chains and square lattices are shown for
the nearest-neighbor tight-binding model with $t_{i,j}=t\delta _{i,j}$ in
Figs. 1(a) and (c), and for the case of long-range overlap integrals in
Figs. 1(b) and (d). Characteristic van-Hove singularities are observed, in
one dimension at the band edges, and in two dimensions at the band center.
For the long-range model, the particle-hole symmetry is broken because of
frustration introduced by the competing overlaps, leading to asymmetries in
the density of states. While the system sizes in Fig. 1 are chosen to be
rather large in order to make contact with the familiar thermodynamic limit,
there are still some visible finite-size remnants, i.e. a faint pole
structure due to the discreteness of the system. These features become much
more pronounced for the few-atom clusters that are studied in the next
sections of this paper.

\section{Target functions and convergence criterion}

While ``nature" gives us densities of states that are constrained by the
dimensionality and symmetry of the underlying lattice, our objective is to
engineer spectral responses with specific shapes that are useful in
designing nanoscale devices. For example, we may wish to produce a
quasi-two-dimensional spectrum in a one-dimensional system or to concentrate
spectral weight in particular energy windows. These goals are achieved by
placing the atomic constituents into optimized symmetry-breaking
configurations which are determined by numerical searches.

\begin{figure}[h]
\includegraphics[width=8.5cm]{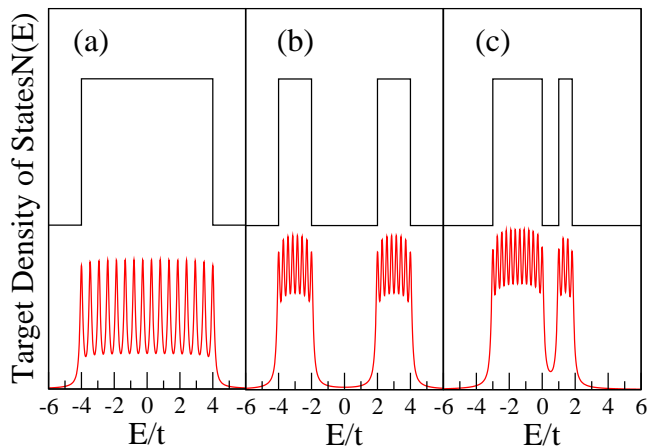}
\caption{ Target densities of states used in this work: (a) top hat
function, centered at E = 0, (b) particle-hole symmetric two-peak function,
and (c) asymmetric two-peak function. For systems with finite numbers of
particles these shapes are approximated by quasiparticle peaks. }
\end{figure}

The specific target functions that are studied in this work are shown in
Fig. 2. They are (a) a flat top hat density of states centered at $E = 0$,
\begin{eqnarray}
N(E)_{target} = \theta(E - E_c) \theta(E + E_c),
\end{eqnarray}
where $\theta (x)$ is the Heavyside function, and $E_c$ is an energy
cut-off, (b) a symmetric two-peak function, centered at $E = 0$,
\begin{eqnarray}
N(E)_{target} & = & \theta(E - E_{c2}) \theta(E + E_{c1})  \nonumber \\
& + & \theta(E - E_{c1}) \theta(E + E_{c2}),
\end{eqnarray}
and (c) a particle-hole-symmetry-breaking function with two unequal peaks,
i.e. more spectral weight on the quasi-hole side ($E < 0$) than on the
quasi-electron side ($E > 0$) of the spectrum,
\begin{eqnarray}
N(E)_{target} & = & \theta(E - E_{c2}) \theta(E + E_{c1})  \nonumber \\
& + & \theta(E - E_{c4}) \theta(E + E_{c3}).
\end{eqnarray}

In systems with finite numbers of tight-binding atoms, these continuous
shapes are approximated by equally spaced poles within the energy windows
where $N(E)_{target}$ is a non-vanishing constant. Here, the delta-functions
are given a finite width of 0.02t. As more atoms are added to the system,
these peaks merge together, approaching the bulk result. For other non-flat
target functions the quasiparticle peak spacing can be varied, e.g.
following a Gaussian or Lorentzian shape. Naturally, not all targets can be
achieved equally well. Factors that influence the achievable match to a
target include the number of available number of atoms and, as will be
shown, a continuous or discrete number of accessible spatial positions. The
dimensionality of the system poses additional constraints. In particular,
there are less available configurations in lower dimensions. This study
focuses on three prototype spectral responses which are targeted by
numerically optimizing configurations of clusters with up to 48 atoms in two
spatial dimensions.

The optimization algorithms seek to minimize the deviation from a given
target density of states, defined by the error function
\begin{eqnarray}
\Delta = \int_{-\infty }^{\infty } dE [N(E) - N(E)_{target}]^2
\end{eqnarray}
which is the least-square difference between the system response
for a given configuration and the target response function. We
have explored a number of numerical techniques, including the
Newton-Raphson steepest descent method, simple downhill random
walk, simulated and triggered annealing, and genetic algorithms.
The advantages and disadvantages of these techniques are briefly
discussed in the appendix. In general, it is found that hybrids of
these methods tend to work best. In the next section, we focus on
adaptive design of atomic clusters in continuous configuration
space without any restrictions to underlying discrete lattices. In
this case, the search space is infinite which generally allows
better convergence to a given target response than in finite
configuration space. However, some experimental realizations of
such structures require deposition of atoms on substrates with
discrete lattice structures.\cite{nazin,nilius,wallis}
Therefore, this case is addressed
separately in the subsequent section.

\section{Atoms-up design of tight-binding clusters in continuous
configuration space}

The first target density of states we would like to study in
detail is the particle-hole symmetric top hat function, centered
at energy E=0, shown in Fig. 2(a). This function represents a
constant density of states for a bulk solid between the energy
cut-offs $\pm E_c$, giving a bandwidth $W = 2 E_c$. Here, we
choose $E_c = 3t$. However, it should be noted that with the
adaptive quantum design approach we are not restricted to this
choice, and target densities of states with quite different
bandwidths can be matched, although often to a lesser degree of
accuracy.

\begin{figure}[h]
\includegraphics[width=15.cm]{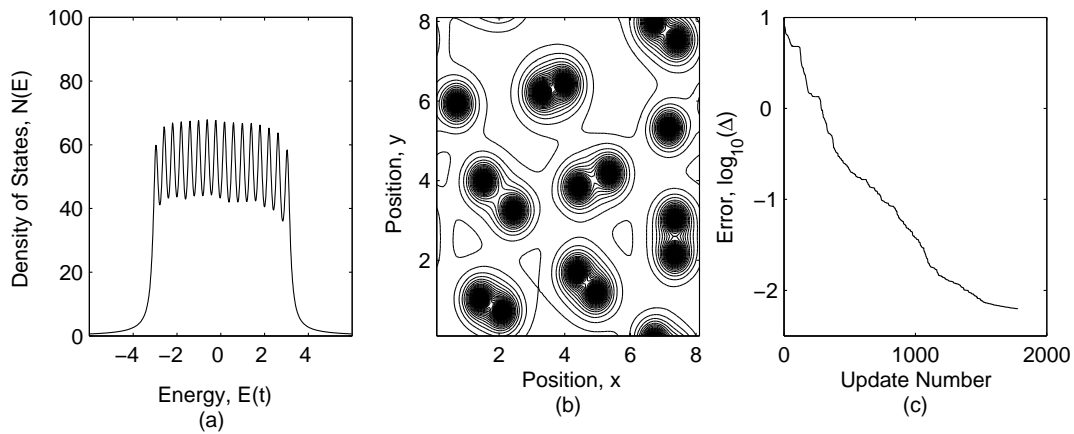}
\caption{ Adaptive quantum design applied to 16 atoms in a two-dimensional
box with periodic boundary conditions. The target density of states is a
symmetric top hat function with bandwidth $6t$. The atomic configurations
are optimized by applying a guided random walk algorithm to the long-range
tight-binding model. (a) The best matching solution. (b) Contour plot of the
potential of the resulting spatial configuration. (c) Convergence to the
target function, $\log _{10}(\Delta )$, with the number of updates. }
\end{figure}

In Fig. 3, the solution for a system with 16 atoms confined to a
box with periodic boundary conditions is shown. The guided random
walk method is applied to optimize the configuration of atoms by
iterative local updates of their positions in order to match the
top hat density of states. As observed in Figs. 3(a) and (c), good
convergence to the target function can be achieved for this case
after less than 2000 updates. A contour plot of the potential
$\sum_{ij} -t_{ij}/ |\vec{r}_i - \vec{r}_j |^{\alpha}$ for the
resulting spatial configuration is shown in Fig. 3(b). Here,
equipotential lines are used to denote the positions and
overlapping wave functions of the atomic constituents in the
system. For this target function one discovers the formation of
dimers with a wide range of inter- and intra-dimer spacings. These
self-assembled building blocks have variable directional
orientation, and are closely packed.

\begin{figure}[h]
\includegraphics[width=15.cm]{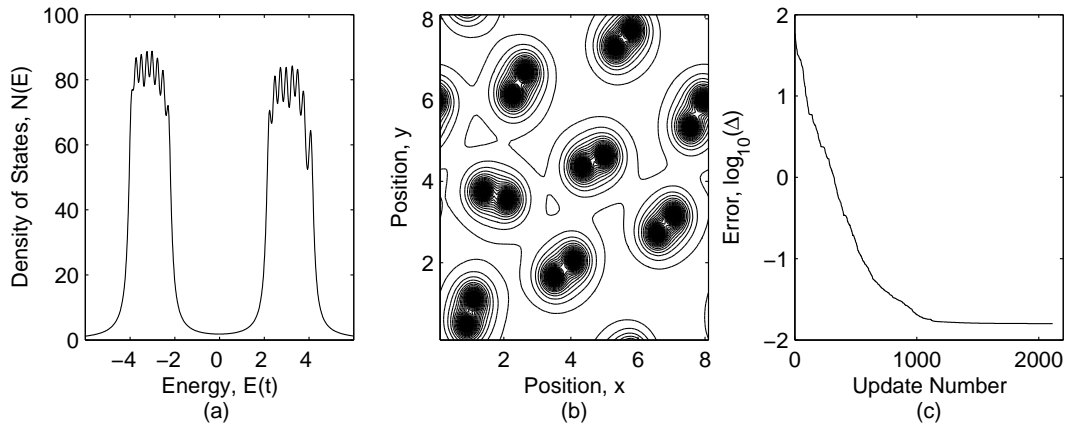}
\caption{ Same as Fig. 3, but for a symmetric two-peak target function with
peaks of bandwidth 2t centered at -3t and 3t. }
\end{figure}

To explore the generic aspects of this result, let us now turn to
the two-peak target function shown in Fig. 2(b). This density of
states has a gap in the center of the spectrum, as may be desired
for the construction of two-state systems such as q-bits. As shown
in Figs. 4(a) and (c), the convergence to this target function is
not quite as good as for the top hat function, but saturation
occurs already at half the number of iterations compared to the
previous example. Interestingly, the optimized spatial
configuration that is found in Fig. 4(b) also displays a
preference for dimer formation.

\begin{figure}[h]
\includegraphics[width=9.cm,angle=270]{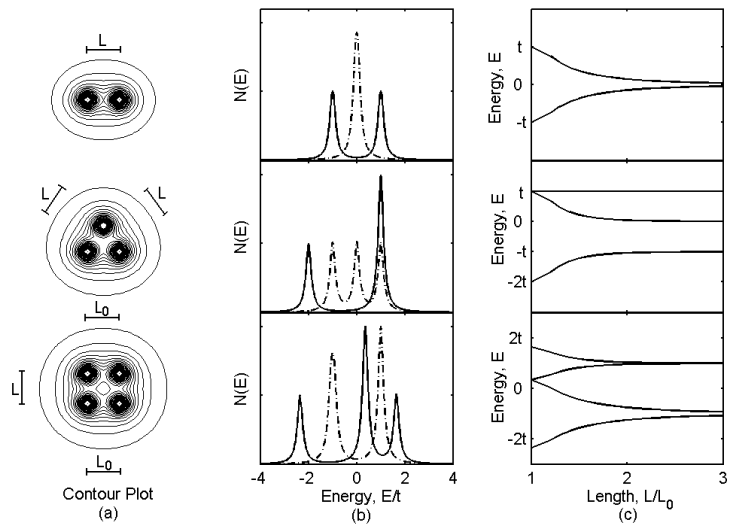} \vspace{-0.2cm}
\caption{ Spectral response of tight-binding dimers, trimers, and
quadrumers. (a) Potential contour plot of dimer, trimer, and
quandrumer of size L$_{\text{0}}$. (b) In the dimer, the
single-atom level (dashed line) is split symmetrically (solid
line). Competing interactions (frustration) in the trimer and
quadrumer lead to asymmetric densities of states. The strength of
the frustration depends on the separation L. (c) Spectral peak
positions as a function of normalized distance L/L$_{\text{0}}$
for dimer, trimer, and quadrumer. The spectra for L/L$_{\text{0}}$
= 1 are shown as solid lines and L/L$_{\text{0}}$ = 3 as 
dashed lines in (b).}
\end{figure}

In the long-range interacting systems we are studying it is, at
least at first sight, not obvious why these target functions
prefer dimer building blocks. Let us address this question by
examining the individual spectra of the molecular building blocks
shown in Fig 5. Each dimer molecule contributes to the
density of states a positive and negative pole with energies $E = \pm t_{12}$%
, where $t_{12}$ is the hopping integral between the two
participating atoms. The zero-energy quasiparticle peaks of two
isolated atoms are split into bonding and anti-bonding
combinations once they form dimer molecules. Therefore, isolated
dimers are ideal building blocks for particle-hole symmetric
densities of states, such as the top hat and the two-peak target
function. The intra-dimer spacing determines the positions of the
$E = \pm t_{12}$ poles via Eq. 2. With an appropriate distribution
of these distances the full target spectrum of the top hat
function can be covered. The poles close to the band center ($E =
0$) are provided by the less tightly bound dimers. For the
two-peak target function the dimer building blocks required to
realize the target spectrum are more tightly bound, and the
intra-dimer spacings need to vary less to achieve this target.

The idea that dimers can be used as building blocks for the
particle-hole symmetric target functions strictly applies only
to isolated
dimers, i.e. the dilute limit, or when potential gradients across
dimers pairs from the presence of adjacent atoms do not break
particle-hole symmetry. The absence of such gradients, even for
relatively high atom densities in a long-range interacting system,
accounts for the success of dimers in satisfying the target
function.

Lower symmetry building blocks, such as trimers and quadrumers shown in Fig.
5, can achieve more complex target functions, in particular those with
broken particle-hole symmetry. Due to frustration, trimer molecules
intrinsically have asymmetric densities of states with unequal spectral
weights on the electron and the hole side of their spectra. Quadrumers have
a symmetric spectral response in the absence of longer-range frustrating
interactions. As an example of how these building blocks enable more complex
target functions, let us consider the asymmetric two-peak density of states
given by Eq. 5 and shown in Fig. 2(c), which has a narrow upper peak and a
wider lower peak, separated by a gap. In Fig. 6 it is observed that an
approximate match can be achieved by the adaptive method. As expected, the
building blocks for this particle-hole-asymmetric target function are
combinations of dimers, trimers, and quadrumers, which partially recombine
into larger clusters. Obviously, in order to achieve asymmetric target
functions more complex building blocks are required. Especially for systems
which contain only a small number of atoms it is important whether the
required building blocks are available, and whether there are
non-participating unbound atoms that may deteriorate convergence and match
to the target function.

\begin{figure}[h]
\includegraphics[width=15.cm]{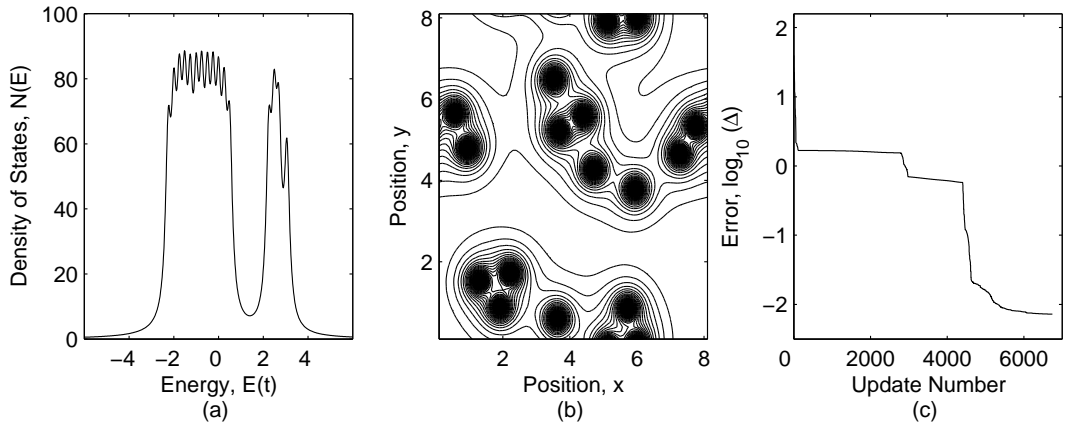}
\caption{ Asymmetric two-peak quasiparticle density of states. The solution
contains dimer, trimer, and quadrumer building blocks. }
\end{figure}

Let us finish this section by addressing the convergence properties of the
three cases that were discussed. So far, only the simplest guided random
walk optimization method was considered in which every \textquotedblleft
downhill step" is accepted. These \textquotedblleft downhill steps" are
local updates of individual atomic position that lead to a better match of
the spectrum to the target density of states. Especially for the most
symmetric target function, this algorithm converges efficiently, with a
small remaining error. For the symmetric two-peak function, convergence
occurs even faster because the required dimer building blocks are more
uniform than for the first case. However, the remaining error is slightly
larger, indicating that this may be a metastable solution which could be
improved by global updates in which whole subclusters are simultaneously
updated. Finally, the convergence plot for the asymmetric two-peak target
function (Fig. 6(c)) shows several plateaus, indicating metastable
configurations that exhibit a high resistance against local updates. For
this case, a much larger number of local updates is required to achieve an
acceptable match. Hence, more complex target functions clearly call for more
sophisticated numerical search tools, including annealing steps,
parallelization, and global updating schemes when available.

\section{Adaptive design in discrete configuration space}

In the previous section local updates were considered which allow atomic
positions to change continuously within a given radius. However, for the
case of atoms deposited on a substrate with a given lattice structure, the
set of available positions is usually discrete, although it may be very
large. This has significant consequences for the adaptive design approach.
The search space of solutions is finite in this case, which makes it
feasible to study more atoms with similar computational effort compared to
the continuous case. At the same time, the discreteness of the lattice can
prohibit favorable configurations that are available in the continuous case,
thus deteriorating convergence properties of the optimization procedure. In
practice, we find that the feasibility of computations for larger numbers of
atoms in the discrete case helps to achieve better matches, as long as the
lattice spacings remain sufficiently small. In this section, we explore the
effects of an underlying grid on the adaptive design procedure. Also, more
advanced techniques are implemented for the numerical optimization,
including hybrids of the genetic algorithm, simulated annealing, and the
guided random walk method.

\begin{figure}[h]
\includegraphics[width=12.cm]{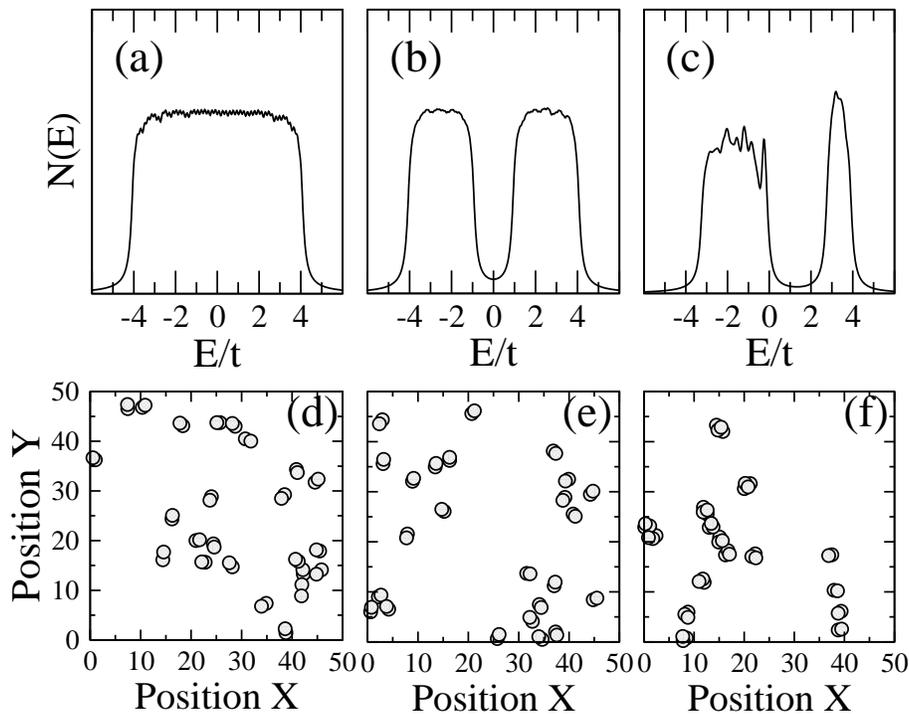}
\caption{ Adaptive design of clusters with 48 atoms on a discrete lattice.
The (a) flat, (b) symmetric two-peak, and (c) asymmetric two-peak densities
of states are the same as discussed in the previous section. The
corresponding atomic configurations are shown in (d), (e), and (f). }
\end{figure}

In Fig. 7, adaptive design results are shown for systems with 48 atoms on a
square lattice with spacing 0.01. The length scale is set by the linear size
of the two-dimensional box (48 $\times $ 48) to which the particles are
confined. These systems are in the dilute limit, and hence the convergence
to the target functions, chosen to be the same as in the previous section,
is good. Also, because of the larger number of particles, there are less
finite size effects. For the top hat target function one obtains a small
matching error of $\Delta = $ 0.000643, for the symmetric two-peak function
one finds $\Delta = $ 0.000605, and for the asymmetric two-peak function the
error is $\Delta = $ 0.150347. Thus, analogous to the continuous case, the
more symmetric targets are easier to achieve. Again, clustering into dimers
is observed for the particle-hole symmetric target functions, whereas
trimers are the preferred building blocks for the asymmetric target. Since
atomic densities in these examples were chosen to be in the dilute limit,
boundary effects and interactions between the building blocks are small. The
resulting configurations have the character of liquids, governed by weak
interactions between the molecular building blocks, and relatively strong
confining forces that lead to the formation of dimers and trimers.

\begin{figure}[h]
\includegraphics[width=8.cm]{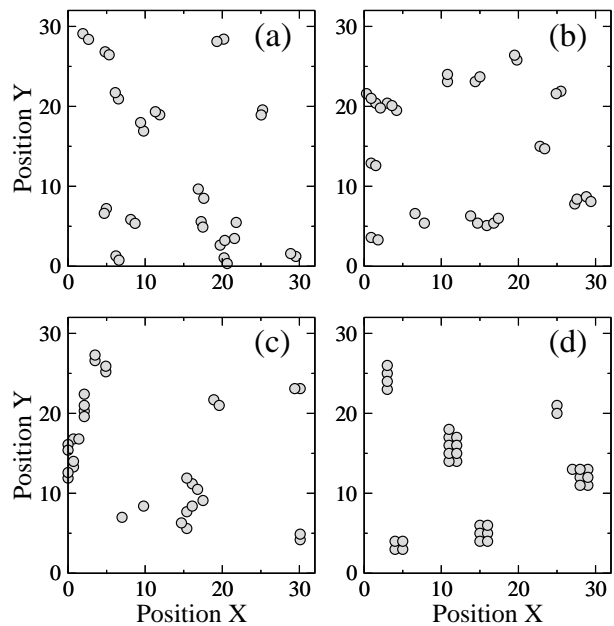}
\caption{ Effect of the coarseness of the underlying lattice on the
formation of building blocks. 32 atoms are confined to a 32 $\times $ 32
square box. The target function is the top hat density of states. The grid
spacing is varied: (a) 0.01, (b) 0.3, (c) 0.7, and (d) 1.0. }
\end{figure}

Next, let us explore the dependence of these solutions on the coarseness of
the underlying lattice. In Fig. 8, the grid spacing is varied over two
orders of magnitude from 0.01 up to 1.0. As expected, the convergence to the
target top hat function deteriorates dramatically as the substrate is made
coarser. For the smallest spacing of 0.01 one converges to a final error of $%
\Delta = $ 0.28457, for a spacing of 0.3 the error is $\Delta = $ 2.45047,
for a spacing of 0.7 the error becomes $\Delta = $ 3.81779, and for the
coarsest case with spacing 1.0 the error is $\Delta = $ 36.3721, indicating
failure of convergence. The corresponding configurations in Fig. 8 show a
strong dependence of the clustering sizes as the system is trying to cope
with less available phase space to match the target function. For the finest
grid spacing (Fig. 8(a)) one observes almost entirely isolated dimers. As
the spacing is increased (Fig. 8(b)), a few small strings and groups are
formed. At grid spacing 0.7 (Fig. 8(c)), the solution is made up mostly of
long strings and dimers in close proximity to each other. Ultimately, for
the coarsest grid spacing of 1.0, (Fig. 8(d)), the final configuration
consists of square and rectangular blocks of atoms. This result demonstrates
that there is a hierarchy of building blocks. The most suitable building
blocks for the given target function are dimers. When these become less
available due to lattice constraints, the adaptive method selects higher
order solutions, i.e. larger size clusters, in order to cope with the more
restricted phase space of possible configurations. Hence, for each level of
coarseness, adaptive design discovers solutions that enable - up to a given
degree of accuracy - a targeted system response.

\begin{figure}[h]
\includegraphics[width=8.cm]{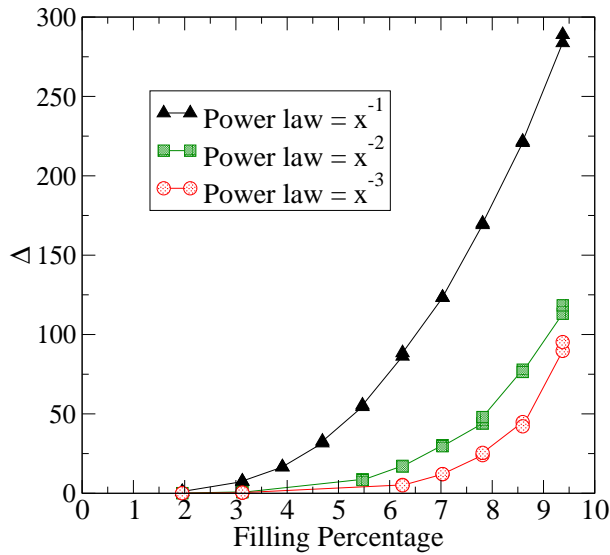}
\caption{ Effect of the density of atoms and the power-law dependence of the
tight-binding overlap integral the convergence of adaptive quantum design. }
\end{figure}

By widening the grid spacings and keeping the box size constant, the density
of atoms in the system is effectively increased, and simultaneously the
available energy levels are spaced further apart. Let us examine these
dependencies by varying the power-law governing the atomic overlaps (Eq. 2),
and by increasing the number of atoms, i.e. the filling fraction, on a fixed
lattice. Results for the achieved convergence are shown in Fig. 9. These
demonstrate that excellent target matches can be achieved in the dilute
limit with filling densities of a few percent. For larger coverages, the
numerical search becomes exponentially less effective, indicating increasing
frustration effects that need to be addressed by global updating schemes.
Higher power-laws imply effective shorter-range atomic overlaps, thus
rendering the system more dilute. This is reflected in Fig. 9, where the
departure from the regime of negligible matching errors is pushed toward
higher filling percentages as $\alpha $ in Eq. 2 is increased.

Some aspects of the long-range tight-binding model on a substrate have
already been confirmed experimentally using scanning tunneling microscopy
(STM) to precisely position gold atoms on the surface of a nickel-aluminum
crystal. In a recent study performed at UCI by Dr. Ho's group,\cite{nilius}
STM measurements show that the splitting in the value of eigenenergies for
Au dimers on NiAl depends inversely on Au atom separation corresponding to $%
\alpha =1$. Here, the grid spacings are dictated by the 0.29 nm
lattice periodicity of available add-atom sites on the NiAl
substrate. Remarkably, the power-law dependence of the effective
overlaps $t_{ij}$ between the deposited atoms takes into account
interactions with the substrate which are typically difficult to
model by first principle computations.

\section{Conclusions}

In this work, adaptive quantum design techniques were applied to tailor the
quasiparticle density of states of atomic clusters, modeled by the
long-range tight-binding Hamiltonian. Broken-symmetry spatial configurations
of atoms were optimized to match target spectra. By applying adaptive search
algorithms, it was shown that matches to target responses can be achieved by
forming hierarchies of molecular building blocks that depend on system
constraints. For example, symmetric top hat and two-peak target densities of
states can be achieved by forming lattices of weakly interacting dimers.
While these are the elementary building blocks for particle-hole symmetric
case target functions, more complex molecules, such as trimer and
quadrumers, are found to dominate the solutions for asymmetric target
functions. Implementation of this approach on discrete lattices,
corresponding to finite configuration space, introduces 
frustration effects that distabilize the elementary building blocks in 
the limit of coarse grid spacings.

The core task of adaptive quantum design is the numerical search for global
minima in typically shallow landscapes of configurations with many local
minima. Since this procedure typically requires many function calls, an
efficient implementation on parallel computers is necessary. For the
purposes of this study, the complexity of the physical model was minimized
in order to limit the computational expense. While the long-range
tight-binding model can be viewed as a semi-realistic testing ground for
adaptive quantum design techniques, it is crucial to apply these algorithm
to more sophisticated models that include, among other ingredients, orbital
directionality, spin degrees of freedom, and electronic correlations.
Furthermore, adaptive design is applicable to related areas in
nanotechnology, including the design of electro-optic components\cite{yu}
and RF systems\cite{gheorma}.

Our enthusiasm for adaptive quantum design is, in part, driven by the
potential for scientific and technological discovery built into the
methodology. We find it appealing to perform machine-based searches for new
configurations, new components, and new sub-systems, as we believe these
will inevitably create new understanding, heuristics, and intuition. The
fact that searches are performed in an ultra-large configuration space
virtually guarantees the discovery of completely new designs and a much more
thorough exploration of the possibilities and capabilities of nanotechnology.

\section{Acknowledgements}

We are grateful to Ioan Gheorma, Weifei Li, Peter Littlewood, Omid Nohadani,
Tommaso Roscilde, and Rong Yu for discussions. We acknowledge financial
support by DARPA and the Department of Energy, Grant No. DE-FG03-01ER45908.
Computational support was provided by the USC Center for High Performance
Computing and Communications and by the NERSCC.

\section{Appendix: Optimization and search algorithms}

The optimization algorithms used in this work are based on local updates of
atomic positions in order to minimize the error $\Delta $ defined in Eq 6.
Each atom in the system is visited periodically, and a trial change of its
position is attempted. Depending on the response in $\Delta $ and on the
specific algorithm, this trial step is either accepted or rejected. For the
case of continuous configuration spaces, these local updates are random
shifts of positions within a given radius. A hard core constraint is
implemented which forbids atoms to be placed on top of each other. For
discrete configuration space, a stochastic distribution function is used to
decide which sites in the neighborhood of the original position of an atom
should be visited in a trial step. While this function is naturally peaked
at nearest-neighbor sites, it has to include a finite probability of longer
range updates in order to avoid getting stuck in local minima of search
space. The particular results discussed in this appendix are obtained for
the continuous case.

\begin{figure}[h]
\includegraphics[width=8.5cm]{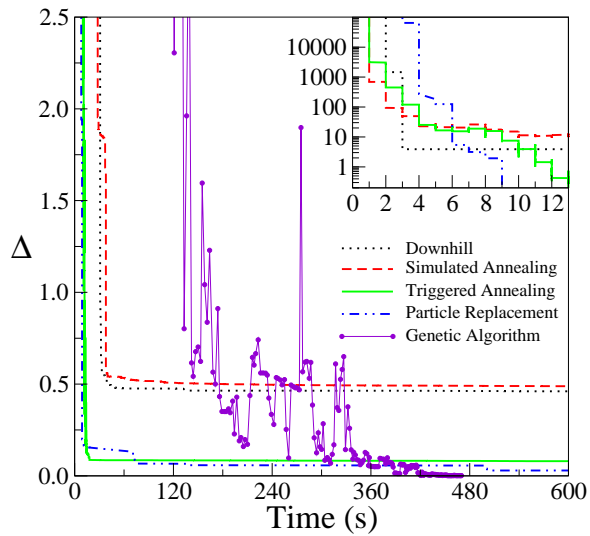} \vspace{-0.5cm}
\caption{ Comparison of the convergence of several optimization algorithms.
The error $\Delta $ is shown as a function of computer run time. The inset
uses a logarithmic scale for $\Delta $. }
\end{figure}

The Newton-Raphson and Broydn methods are multi-dimensional generalizations
of the one-dimensional secant method\cite{press}. Unfortunately, their
global convergence is rather poor for more than 20 variable parameters.
Therefore, they are only applicable to the smallest system sizes, and are
not useful for the non-linear multi-parameter searches required for adaptive
quantum design.

In the guided random walk or ``downhill method", each random step that
results in a smaller target error $\Delta$ is accepted.\cite{yu} Random
steps are trial spatial variations about a particle's position. This guided
random walk technique is a quickly implemented power horse. However,
especially for shallow landscapes of solutions it gets easily stuck in local
minima.

Simulated annealing uses an effective temperature representing the
likelihood of accepting a step that does not minimize the function. This
temperature is lowered slowly - at a rate of 10\% - with each iteration. The
initial temperature is taken as $T_{init} = 5t$. This method is better at
avoiding local minima than the previous techniques. However, it takes a
relatively long time to converge.

The triggered annealing method is a hybrid of the downhill and the simulated
annealing method. It implements the downhill method until minimizing steps
become hard to find, at which point the simulated annealing method is used
to escape from local minima. Parameters are chosen the same as for the
simulated annealing method. Triggered annealing tends to converge relatively
quickly if there are only a few local minima.

The particle replacement method uses the simple downhill method for guided
random updates. In addition, it identifies particles that have not been
updated for an extended period, because of being stuck in a local minimum,
and assigns them to a new random position within the lattice boundaries. In
our implementation, a particle is replaced if there are 10 idle iterations
without a successful downhill update for that particular particle. Note that
this particle replacement update is different from random step updates
because it is independent of the previous position of the particle.

In genetic algorithms a population of possible solutions is created. Those
that best minimize the function are allowed to take part in creating a new
generation of possible solutions.\cite{whitley} These methods are generally
good at avoiding local minima, and are also easily implemented on parallel
computers. They typically require more function calls than other search
algorithms. 

In order to illustrate the efficiency of these various approaches each
method is used to match a flat top hat target function on a one-dimensional
lattice with 24 tight-binding atoms and a box size of 96. Particles exiting
the box on one end, enter it from the other side via periodic boundary
conditions. The target density of states
is chosen to be the symmetric top-hat function, with poles evenly spaced
between $E=-4t$ and $4t$. All of these benchmark runs are started with atoms
placed randomly along a chain. The computer is a Pentium III 1 GHz with
2GB pc133 memory, and the additional seven nodes used by the genetic
algorithm method are Pentium III 850 MHz processors with 1GB memory. Each
method is run for 600 seconds.

As shown in Fig. 10, all methods, with the exception of the genetic
algorithm, converge rapidly within the first minute of run time, and show
only relatively small corrections afterwards. The two techniques with the
fastest convergence (inset of Fig. 10) are the downhill and the simulated
anealing methods. However, their asymptotic error functions remain
relatively large, indicating that they get easily stuck in local minima. In
contrast, the triggered annealing and particle replacement methods yield
much better matches to the target function, while still converging
relatively fast. As shown in the inset, the annealing methods sometimes
accept trial steps in the ``wrong" direction in order to avoid local minima.
Finally, the genetic algorithm takes a relatively long time to converge.
However, it yields by far the best match to the target density of states
after about 7 minutes run time. In order to ensure best matches to the
target, this last method is therefore used whenever the computational effort
allows it.


\begin{thebibliography}{9}

\bibitem{harrison} W. A. Harrison, ``Electronic Structure and the Properties
of Solids", W.H. Freeman and Company (1980).

\bibitem{nazin} G.V. Nazin, X.H. Oiu, and W. Ho, Science \textbf{308}, 77
(2003); Phys. Rev. Lett. \textbf{90}, 216110 (2003).

\bibitem{nilius} N. Nilius, T.M. Wallis, M. Persson, and W. Ho, Phys. Rev.
Lett. \textbf{90}, 196103 (2003); N. Nilius, T.M. Wallis, and W. Ho, Science
\textbf{297}, 1853 (2002).

\bibitem{wallis} T.M. Wallis, N. Nilius, and W. Ho, Phys. Rev. Lett. \textbf{%
89} 236802 (2002).

\bibitem{press} W.H. Press, S.A. Teukolsky, W.T. Vetterling, and B.P.
Flammery, ``Numerical Recipes: The Art of Scientific Computing", Cambridge
(1986).

\bibitem{yu} Y. Chen, R. Yu, W. Li, O. Nohadani, S. Haas, and A.F.J. Levi,
J. Appl. Phys. \textbf{94}, 6065 (2003).

\bibitem{whitley} see e.g. D. Whitley, Statistics and Computing \textbf{4},
65 (1994).

\bibitem{gheorma} I. L. Gheorma, S. Haas, and A. F. J. Levi, cond-mat/0310212,
to be published in J. Appl. Phys (2004).

\end{thebibliography}
\end{document}